\documentclass[12pt]{article}
\usepackage{amsmath,amssymb,amsthm,amsxtra,overpic,bbm,bm,epsfig,ulem,multirow}
\usepackage{color,cite}
\usepackage{epstopdf}

\def\thefootnote{\fnsymbol{footnote}}

\usepackage{array}

\begin{document}

\vspace{0.2cm}

\begin{center}

{\Large\bf New Realization of the Conversion Calculation for Reactor Antineutrino Fluxes} 
\end{center}

\vspace{0.2cm}

\begin{center}
{\bf Yu-Feng Li$^{a}$}\footnote{Email: liyufeng@ihep.ac.cn}
and {\bf Di Zhang$^{a}$}\footnote{Email: zhangdi@ihep.ac.cn} \\
{$^a$Institute of High Energy Physics, and School of Physical
Sciences, University of Chinese Academy of Sciences, Beijing 100049, China}
\end{center}

\vspace{1.5cm}

\begin{abstract}
Validation of the effective conversion method in the reactor antineutrino flux calculation is examined using the \textit{ab initio} calculation of
the electron and antineutrino spectra from the state-of-the-art nuclear database.
It turns out that neglecting the shape variation of beta decay branches between the allowed and forbidden transitions would induce significant
bias in the total inverse-beta-decay yields and energy spectral distributions. We propose a new realization of the conversion method with both the allowed and forbidden virtual branches, and apply it to both the
\textit{simulated} data from the nuclear database and
\textit{real} data from the fission measurements at ILL by virtue of statistical properties of the allowed and forbidden decays in the database.
Two kinds of dominant uncertainty sources are identified and it turns out that
the new realization of the conversion calculation can largely reduce the rate and spectral bias and thus present a reliable prediction of the antineutrino fluxes
if accurate beta decay information is available in the high endpoint energy range.
\end{abstract}

\begin{flushleft}
\hspace{0.8cm} PACS number(s): 14.60.Lm, 23.40.-s, 25.85.Ec.
\end{flushleft}

\def\thefootnote{\arabic{footnote}}
\setcounter{footnote}{0}

\newpage

\section{Introduction}

Electron antineutrinos from nuclear reactors have played important roles in the history of neutrino physics
for studying fundamental properties of neutrinos
(see the reviews in Refs.~\cite{Bemporad:2001qy,Vogel:2015wua,Qian:2018wid}). The reactor
antineutrinos are produced in the beta decays of fission fragments associated with four main fissionable isotopes
${}^{235}\text{U}$,
${}^{238}\text{U}$,
${}^{239}\text{Pu}$, and
${}^{241}\text{Pu}$.
There are two methods to obtain a direct calculation of the antineutrino fluxes from these isotopes.
One can either employ the \textit{ab initio} method~\cite{Davis:1979gg,Avignone:1980qg,Klapdor:1982zz,Klapdor:1982sf,Mueller:2011nm,Fallot:2012jv,Dwyer:2014eka,Hayes:2015yka,Yoshida:2018zga}
by a direct summation of the antineutrino energy spectrum in the beta decay of each fission fragment
from the state-of-the-art nuclear database, or apply the effective method with a conversion
procedure~\cite{VonFeilitzsch:1982jw,Schreckenbach:1985ep,Hahn:1989zr,Huber:2011wv,Haag:2013raa}
 based on the measurement of the integral electron energy spectrum for the main fissionable isotopes.
However, the reliability and accuracy of the isotopic flux calculations should be carefully examined in order to
resolve the reactor antineutrino rate and spectral anomalies~\cite{Mention:2011rk,Giunti:2019qlt,Seo:2014xei,An:2015nua,Abe:2015rcp,Ko:2016owz}
and serve for the neutrino mass hierarchy measurement in future reactor antineutrino experiments~\cite{An:2015jdp}.

There are more than 6000 beta decay transitions of the fission fragments contributing to each of the four fissionable isotopes
${}^{235}\text{U}$,
${}^{238}\text{U}$,
${}^{239}\text{Pu}$, and
${}^{241}\text{Pu}$.
In the \textit{ab initio} summation method, the fission yields and the beta decay information of fission fragments are the
prerequisite to calculate the isotopic neutrino fluxes. Therefore, the accuracy of the calculation resides in the uncertainties in the
fission yield and beta decay information. The fission yields have been evaluated by different international nuclear databases,
but large uncertainties and even their incompleteness for many important fragments are still a problem for the current nuclear
databases. As for the beta decay information, the initial and final state quantum numbers as well as their decaying branching ratios
are needed to characterize the beta decay spectra and calculate the antineutrino flux of each fission fragment, which, however,
are not always known for all the fragments.

To overcome the aforementioned problems of the \textit{ab initio} method, an effective conversion
method has been developed thanks to the measurements of aggregate electron spectra associated with the thermal neutron induced fission of
${}^{235}\text{U}$, ${}^{239}\text{Pu}$, and
${}^{241}\text{Pu}$ at ILL, Grenoble, France in 1980¡¯s~\cite{VonFeilitzsch:1982jw,Schreckenbach:1985ep,Hahn:1989zr}, and the fast neutron induced fission of ${}^{238}\text{U}$ at
FRMII in Garching, Germany in 2011~\cite{Haag:2013raa}. Since electron antineutrinos and electrons are simultaneously produced in one single beta decay branch,
the isotopic antineutrino fluxes are then obtained with the conversion method by assuming a set of
virtual beta decay branches and fitting the electron spectra of fission isotopes to be consistent with the corresponding measurements.
In this method, the shape characteristics and the relation between the
nuclear charge number and the endpoint energy of the beta decay branches are important uncertainty sources
in the calculations of the isotopic antineutrino fluxes. In previous publications, the reliability and accuracy of the conversion method have
been tested with the assumption of allowed beta decay transitions for all the virtual branches~\cite{Huber:2011wv,Vogel:2007du}.
However, this simple treatment has been challenged by
the incompatibility of the total inverse-beta-decay rate and energy spectrum between the theoretical calculations and experimental
measurements~\cite{Mention:2011rk,Giunti:2019qlt,Seo:2014xei,An:2015nua,Abe:2015rcp,Ko:2016owz}.
Moreover, according to the current nuclear database the beta decay branches of the first forbidden type contribute to around 30\% of the total
branches~\cite{Hayes:2013wra,Hayes:2016qnu} in each fission isotope and they may induce a large
uncertainty~\cite{Hayes:2013wra,Hayes:2016qnu,Fang:2015cma,Stefanik:2017dbz,Hayen:2018uyg}
in the antineutrino flux predictions.

In this work, we plan to make a systematic test for the reliability and accuracy of the effective conversion using the state-of-the-art \textit{ab initio} calculation of
the electron and antineutrino spectra as the reference flux model of electrons and antineutrinos from the fission isotopes. We first show that the systematic
uncertainty of model predictions can be controlled better than 1\% if all the transitions are assumed to be allowed in both the reference model calculation and the conversion procedure.
However, significant bias is observed in the high energy range of the antineutrino spectrum if the shape factor for the forbidden beta decay is considered in
the construction of the reference model but all the virtual branches are taken as allowed in the conversion.
Furthermore, we propose a new realization of the conversion method with both the allowed and forbidden virtual branches, and apply to both the
\textit{simulated} data from the nuclear database and \textit{real} data from the fission measurements at ILL by virtue of the statistical properties of the allowed and forbidden decays in the nuclear database.
Two kinds of dominant uncertainty sources are identified and it turns out that
one could largely reduce the bias of neglecting forbidden transitions and achieve a reliable reactor antineutrino
flux calculation if the types and their ratios of the first forbidden beta decays are accurately known in a statistical way.

This work is organized as follows. In Sec.~2 we review the description of beta decays and the characteristics of the beta decay energy spectrum.
Then we make the validation test of the conversion method using the state-of-the-art nuclear database and propose a new realization of the conversion
calculation in Sec.~3. In Sec.~4 we apply the new realization of the conversion calculation to the ILL $^{235}{\rm U}$ beta spectrum.
Finally the concluding remarks are presented in Sec.~5.

\section{Characteristics of the beta decay spectrum}

Since both the \textit{ab initio} summation method and effective conversion method need an accurate description of the energy spectrum of a single beta decay,
in this section, we plan to review the analytical formulation of the beta decay theory. It is well known that
the electron spectrum shape of one beta decay branch can be expressed as \cite{Fermi:1934hr}
\begin{eqnarray}
N_{\beta}(E_{e}) = K p_e E_e (E_0 - E_e)^2 F_{}(Z,E_e) C(Z,E_e) [1 + \delta(Z,A,E_e)] \;,
\end{eqnarray}
where $K$ is the normalization factor, $E_e$ and $p_e$ are the electron energy and momentum, $E_0$ is the total energy release in the beta decay,
and $E_0-m_{e}$, with $m_{e}$ being the electron mass, is usually defined as the endpoint energy and denotes the maximal energy that the antineutrino can carry in the decay process.
$F_{}(Z,E_e)$ is the Fermi function describing the effect of the Coulomb field on the outgoing electron,
$C(Z,E_e)$ is shape factor accounting for the energy and momentum dependence of nuclear matrix elements,
and $\delta(Z,A,E_e)$ describes the corrections to the spectrum shape including finite size (FS) correction,
weak magnetism (WM) correction and radiative corrections, where $Z$ and $A$ are the charge and nucleon numbers of the daughter nucleus.
The form of shape factor $C(Z,E_e)$ depends on the beta decay transition type. For the allowed transition, $C(Z,E_e) = 1$,
but it is more complex for the forbidden transitions whose expressions are different according to the corresponding transition operators.

For reactor antineutrinos from beta decays of fission fragments, the beta-decaying nuclei have large $Z$ values and thus the Coulomb potentials are very strong. Therefore the energy levels which can undergo the Fermi transitions between the mother and daughter nuclei are much higher compared to the ground levels of mother nuclei. Meanwhile, the GT transitions are energetically possible and will be discussed in the current manuscript. 

The expression of Fermi function is
\begin{eqnarray}
F_{}(Z,E_e) = 2(\gamma+1)(2p_e R)^{2(\gamma-1)} e^{\pi y} \left| \frac{\Gamma(\gamma+iy)}{\Gamma(2\gamma+1)} \right|^2
\end{eqnarray}
with $\gamma = \sqrt{1-(\alpha Z)^2}$ and $y = \alpha Z E_e/p_e$, where $\alpha$ is the fine structure constant,
$R$ is the nuclear radius and is given as a function of $A$, that is~\cite{Elton:1958}
\begin{eqnarray}
R = 1.121 A^{1/3} + 2.426 A^{-1/3} - 6.614 A^{-1},
\end{eqnarray}
in units of $\rm fm$.

The radiative corrections are different for the electron and antineutrino. Therefore, to obtain the antineutrino spectrum,
it not only requires to substitute $E_{\overline{\nu}} = E_0 - E_e$ in Eq.~(1), but also to change the radiative correction from the electron to the antineutrino one.
The radiative corrections to electron and antineutrino spectrum are~\cite{Sirlin:1967zza,Sirlin:2011wg}
\begin{eqnarray}
\delta^{e}_{\rm QED} = \frac{\alpha}{2\pi} g(E_e,E_0) \;, \qquad \delta^{\overline{\nu}}_{\rm QED} = \frac{\alpha}{2\pi} h(E_e,E_0) \;,
\end{eqnarray}
with
\begin{align}
g(E_e,E_0) = & 3\ln{\left( \frac{M_N}{m_e} \right)} - \frac{3}{4} + 4 \left( \frac{\tanh^{-1}{\beta}}{\beta} - 1 \right) \left[\frac{E_0 - E_e}{3E_e} - \frac{3}{2} + \ln{\left( \frac{2(E_0 - E_e)}{m_e} \right)} \right]
\nonumber
\\
& + \frac{4}{\beta} L\left( \frac{2\beta}{1+\beta} \right) + \frac{1}{\beta} \tanh^{-1}{\beta} \left[ 2(1 + \beta^2) + \frac{(E_0 - E_e)^2}{6E^2_e} - 4 \tanh^{-1}{\beta}\right] \;,
\\
h(\hat{E},E_0) =& 3\ln{\left(\frac{M_N}{m_e}\right)} + \frac{23}{4} + \frac{8}{\beta} L \left(\frac{2\hat{\beta}}{1+\hat{\beta}}\right) + 8\left(\frac{\tanh^{-1}{\hat{\beta}}}{\hat{\beta}}-1\right) \ln \left(\frac{2\hat{E} \hat{\beta}}{m_e}\right)
\nonumber
\\
&+ 4\frac{\tanh^{-1}{\hat{\beta}}}{\hat{\beta}} \left(\frac{7 + 3 \hat{\beta}^2}{8} - 2\tanh^{-1}{\hat{\beta}}\right) \;,
\end{align}
where $L(x)$ is defined as $L(x) = \int^{x}_{0} dt/t \ln{(1-t)}$, $\beta = p_e/E_e$, $\hat{E} = E^{}_0 - E^{}_{\overline{\nu}}$, $\hat{\beta} = \hat{p}/\hat{E}$ with $\hat{p} = \sqrt{\hat{E}^2 - m^{2}_{e}}$, $M_N$ and $m_e$ are nucleon and electron mass, respectively.

In the order of $\alpha Z$, the FS correction to the Fermi function of the allowed Gamow-Teller (GT) beta decays is given as~\cite{Hayes:2013wra,Wang:2016rqh}
\begin{eqnarray}
\delta_{\rm FS} = -\frac{3}{2} Z \alpha \left\langle r \right\rangle_{(2)} \left(E_e - \frac{ E_{\nu}}{27} + \frac{m^2_e}{3E_e} \right)
\end{eqnarray}
where $\left\langle r \right\rangle_{(2)} = (36/35) R$, if the weak and charge densities are assumed to be uniformly distributed with the radius of $R$.
Because the FS correction is operator dependent, it is difficult to derive a general and correct expression for all the transitions
and there has not been a satisfactory FS correction for the first forbidden transitions. Thus throughout this work,
the correction for the allowed GT beta decay as in Eq.~(7) will be applied to all transitions, including the forbidden ones.

The WM correction is induced by the interference between the magnetic moment distribution of the vector current
and the spin distribution of the axial current~\cite{Hayes:2016qnu,Wang:2017htp}.
Therefore, it is also true that different transition types have different WM corrections.
We directly take the WM corrections from Ref.~\cite{Hayes:2013wra}, which are listed in the final column of Tab.~1,
where $\mu_{\nu} = 4.7 $ is the nucleon isovector magnetic moment, $M_N$ is the nucleon mass, and $g_A$ is the axial vector coupling constant.

Now we come to the form of the shape factor $C(Z,E_e)$ for forbidden transitions.
In our calculation, all the forbidden transitions are taken as the first forbidden GT transitions and we consider three main and representative first forbidden GT transitions
as listed in the second, third and fourth rows of Tab.~1:
the nonunique first forbidden GT transition with $\Delta J^\pi=0^-$, the nonunique first forbidden GT transition with $\Delta J^\pi=1^-$, and
the unique first forbidden GT transition with $\Delta J^\pi=2^-$.
In Tab.~1, we list two different calculations of the shape factor $C(Z,E_e)$ in the fourth and fifth columns by considering the electron wave function
in the plane wave approximation (PWA) at the nuclear radius~\cite{Hayes:2013wra}
or using the exact relativistic calculation (ERC) of the Dirac wave function~\cite{Stefanik:2017dbz} respectively.
In the latter exact relativistic calculation, the lowest terms of several Fermi-like functions are introduced and can be written as follows~\cite{Stefanik:2017dbz}:
\begin{table}
\centering
	\caption{The shape factors $C(Z,E_e)$ and WM corrections for the allowed and first forbidden GT transitions. The fourth and sixth columns are the shape factor calculated with the plane wave approximation and WM corrections respectively~\cite{Hayes:2013wra}, and the fifth column is the shape factor using the exact relativistic calculation of the Dirac wave function~\cite{Stefanik:2017dbz}.}
	\vspace{0.25cm}
	\renewcommand\arraystretch{1.3}
	\resizebox{\textwidth}{15mm}{
	\begin{tabular}{llllll}
		\hline\hline
		\multirow{2}{*}{Classification}& \multirow{2}{*}{$\Delta J^\pi$}  & \multirow{2}{*}{Operator}& \multicolumn{2}{c}{Shape Factor $C(E_e)$} &
		\multirow{2}{*}{WM correction $\delta_{WM}(E_e)$}
		\\
		\cline{4-5}
		 &  &  & Plane wave approximation &
		Exact relativistic calculation&
		\\
		\hline
		Allowed GT & $1^+$ & $\Sigma \equiv \sigma \tau$ & 1 & 1 & $\frac{2}{3} \frac{\mu_\nu - 1/2}{M_N g_A}(E_e \beta^2 - E_\nu)$
		\\
		Nonunique first forbidden GT & $0^-$ & $ \left[\Sigma,r\right]^{0-}$ & $p^{2}_{e} + E^{2}_{\nu} + 2 \beta^2 E^{}_{\nu} E^{}_{e}$ & $E^{2}_{\nu} + p^{2}_{e} \tilde{F}_{p_{1/2}} + 2 p^{}_{e} E^{}_{\nu} \tilde{F}_{sp_{1/2}} $ & 0
		\\
		Nonunique first forbidden GT & $1^-$ &  $ \left[\Sigma,r\right]^{1-}$ & $p^{2}_{e} + E^{2}_{\nu} - \frac{4}{3} \beta^2 E^{}_{\nu} E^{}_{e}$ & $E^{2}_{\nu} + \frac{2}{3} p^{2}_{e} \tilde{F}_{p_{1/2}} + \frac{1}{3} p^{2}_{e} \tilde{F}_{p_{3/2}} - \frac{4}{3} p^{}_{e} E^{}_{\nu} \tilde{F}_{sp_{1/2}} $ &
		$\frac{\mu_\nu - 1/2}{M_N g_A}\frac{(E_e \beta^2 - E_\nu)(p^{2}_{e} + E^{2}_{\nu}) + 2 \beta^{2} E_{e} E_{\nu} (E_{\nu} - E_{e})/3}{p^{2}_{e} + E^{2}_{\nu} - 4\beta^2 E_{\nu} E_{e}/3}$
		\\
		Unique first forbidden GT & $2^-$ & $ \left[\Sigma,r\right]^{2-}$ & $p^{2}_{e} + E^{2}_{\nu}$ & $E^{2}_{\nu} + p^{2}_{e} \tilde{F}_{p_{3/2}}$ & $\frac{3}{5} \frac{\mu_\nu - 1/2}{M_N g_A}\frac{(E_e \beta^2 - E_\nu)(p^{2}_{e} + E^{2}_{\nu}) + 2 \beta^{2} E_{e} E_{\nu} (E_{\nu} - E_{e})/3}{p^{2}_{e} + E^{2}_{\nu} }$
		\\
		\hline
		\hline
	\end{tabular}}
\end{table}
\begin{eqnarray}
\tilde{F}_{p_{3/2}}(E_e,R) &\simeq& F_1(E,Z)/F_{0}(E,Z) \;,
\nonumber
\\
\tilde{F}_{p_{1/2}}(E_e,R) &\simeq& \left[ \left( \frac{\alpha Z}{2} + \frac{E_e R}{3} \right)^2 + \left( \frac{m_{e} R}{3} \right)^2 - \frac{2 m^2_{e} R}{3E_e} \left( \frac{\alpha Z}{2} +\frac{E_e R}{3} \right) \right]/j^2_1 (p_e R) \;,
\nonumber
\\
\tilde{F}_{sp_{1/2}}(E_e,R) &\simeq& \left[ \left( \frac{\alpha Z}{2} + \frac{E_e R}{3} \right) - \frac{m^2_{e} R}{3 E_e} \right]/\left( j_0(p_e R) j_1(p_e R) \right) \;,
\end{eqnarray}
where $F_{0}(E,Z) = 2/(1 + \gamma) F (E,Z)$~\cite{Wilkinson:1974doj}, and $F_1(E,Z)$ are the Fermi functions defined in Ref.~\cite{Doi:1985dx},
$j_0(p_e R)$ and $j_1(p_e R)$ are the spherical Bessel functions.

\begin{figure}
\centering
\includegraphics[width=0.98\linewidth]{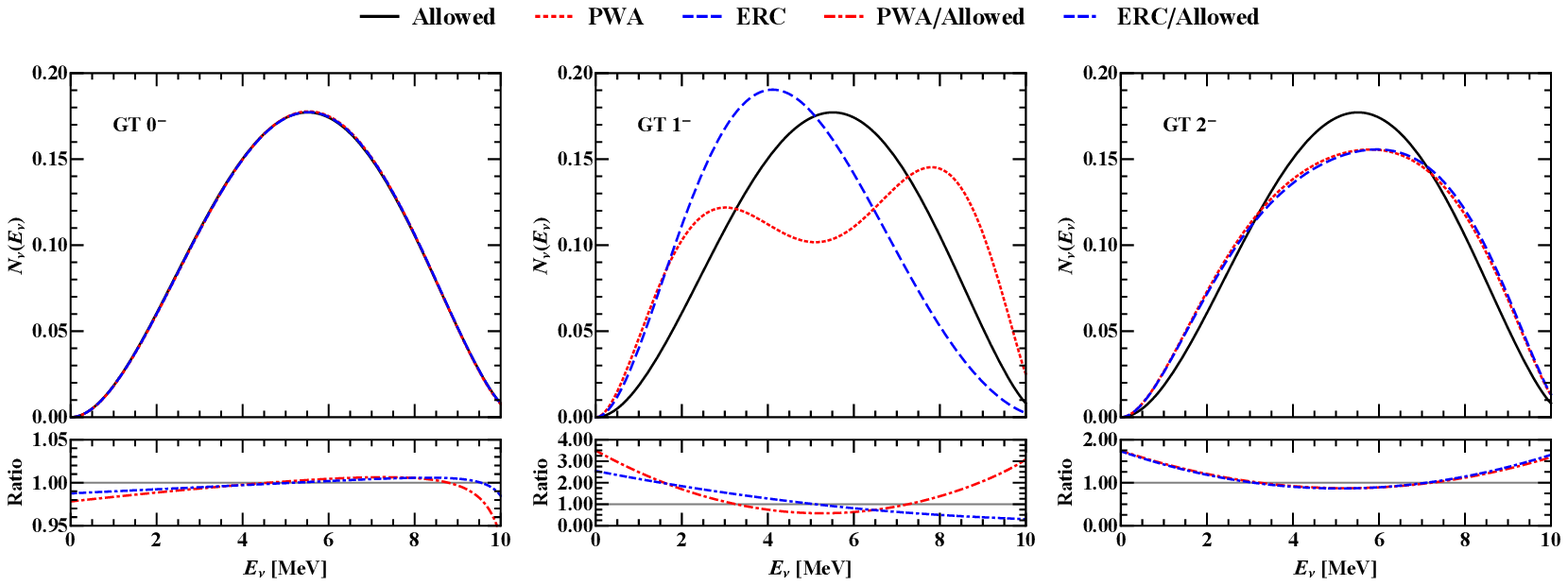}
\caption{The antineutrino spectra with different decay transition types with $Z = 47$, $ A = 117$ for the daughter nucleus and $E_{0}-m_{e}= 10\;\rm{MeV}$.
The left, middle and right panels are for the cases of forbidden GT transitions of $0^-$, $1^-$ and $2^-$ respectively.
The black lines are the shape of allowed GT transitions. 
The red dotted and blue dashed lines stand for the spectra obtained in the plane wave approximation (PWA) and
the exact relativistic calculation (ERC) of the Dirac wave function respectively.
The ratios between the allowed and forbidden spectra are also shown in the
corresponding lower panels.}
	\label{fig:1}
\end{figure}
To show the effect of different shape factors on the beta decay spectrum, we consider a hypothetical transition with $Z = 47$, $ A = 117$ and $E_{0} - m_{e}= 10 \;\rm{MeV}$,
by assuming the allowed GT transition or different types of forbidden transitions.
Comparisons between the antineutrino energy spectra of the allowed and forbidden beta decay are shown in Fig.~1,
where the FS, WM and radiative corrections have also been taken into account.
The ratios between the allowed and forbidden spectra are shown in the
corresponding lower panel below the energy spectrum plots.
From the figure, one can observe that the spectrum of the first forbidden GT $0^-$ transition agrees with that of the allowed transition
at the level of better than 2\%, but those of the first forbidden GT $1^-$ and $2^-$ transitions deviate significantly from the shape of
the allowed beta decay, where the ratios can be as large as a factor of two or three.
One should also be noted that for the first forbidden GT $1^-$ transition there is a large discrepancy between the beta decay spectra of
the plane wave approximation and exact relativistic calculation, where a double peak feature is observed in the case of the plane wave approximation,
which seems to be not realistic. In the following, we shall use
the shape factors $C(Z,E_e)$ of the exact relativistic calculation to describe the beta decay spectra of first forbidden transitions.

The case of ERC can be restored to PWA when one makes the approximation of $\alpha Z \rightarrow0$ and neglects high order terms of $p_{e}R$. Taking the extreme case with $Z = 47$, $ A = 117$ and $E_{0}-m_{e}= 10\;\rm{MeV}$, one can calculate that $\alpha Z\simeq0.34$ and $p_{e}R\simeq0.32$, which turns out to be a large effect and should be carefully included in the shape factor calculations. Regarding the uncertainty of the ERC shape factor, one can anticipate it may reach around 10\% for the extreme case. However since reactor antineutrinos are the summation of all the beta decays, and considering the suppression effects from the fission fraction and branching ratio as well as the smaller nuclear charge $Z$ and endpoint energy, the real uncertainty from the shape factor should be much smaller than 10\%. An actual evaluation should be obtained by the direct uncertainty propagation, and possible correlation should also be carefully treated.

\section{Validation using the nuclear database}

In this section, the validation test for the conversion calculation of the isotopic antineutrino fluxes will be constructed
using the state-of-the-art nuclear database, where the cumulative fission yield data are from the Evaluated Nuclear Data File (ENDF) B.VIII.0,
and the beta decay information are from the database of the Evaluated Nuclear Structure Data Files (ENSDF).
We first employ the \textit{ab initio} summation calculation to generate the reference models of isotopic fluxes for the antineutrino and electron.
The beta decay spectrum described in Eq.~(1) will be used for each branch of the fission fragments.
We generate two groups of reference isotopic flux models for the antineutrino and electron, one assuming all the branches belong to the allowed GT transition,
and the other one considering the shape factor $C(Z,E_e)$ shown in Tab.~1 according to the beta decay property of each branch.
In the conversion method, we use the isotopic electron fluxes as the mock data to carry out the conversion procedure from the electron to antineutrino energy spectra.
Then the converted antineutrino fluxes will be compared with the original isotopic antineutrino fluxes in the reference model in order to test the reliability and accuracy of the conversion method itself.
In this work we shall always present the results of ${}^{235}\text{U}$. The cases for ${}^{238}\text{U}$, ${}^{239}\text{Pu}$, and ${}^{241}\text{Pu}$ have been tested
and the conclusion turns out to be consistent with that of ${}^{235}\text{U}$.

\subsection{Statistical properties of the beta decay database}

In the effective conversion calculation, one usually assume a set of virtual beta decay branches in order to fit with the isotopic electron fluxes.
Since the virtual branch does not correspond to a real beta decay in the nuclear database, typical decay characteristics, including the endpoint
energy, the nuclear charge number, or even the type of forbiddenness, should be assigned to the virtual beta decay according to the statistical properties of
the database.

In the state-of-the-art nuclear database there are more than 6000 beta decay transitions of the fission fragments contributing to the fission isotopes, among which
around 30\% are considered as the forbidden transitions. In the database, the beta decay type can be distinguished as allowed, nonunique forbidden and unique forbidden
transitions. Our selection of the transition type is based on the following strategy. We first treat the transitions with full spin-parity information of both the mother and daughter nuclei. Then if multiple spin-parity information is provided for the final state, we choose the first one as our default selection. Finally when the spin-parity information is not available, we simple group the beta decay decay as the GT allowed one. The discrimination between GT $0^{-}$ and GT $1^{-}$ is expected to be one of the main uncertainty of the conversion method.

Based on the beta decay and fission yield information of ${}^{235}\text{U}$, we present the relative ratios of different beta decay transitions as the function the endpoint energy for
each $1\;{\rm MeV}$ energy interval in Fig.~2, where the relative ratios are calculated using the weighted summation of the fission yield times branching ratio of
each branch in the endpoint energy interval.
\begin{figure}
	\centering\includegraphics[width=0.65\linewidth]{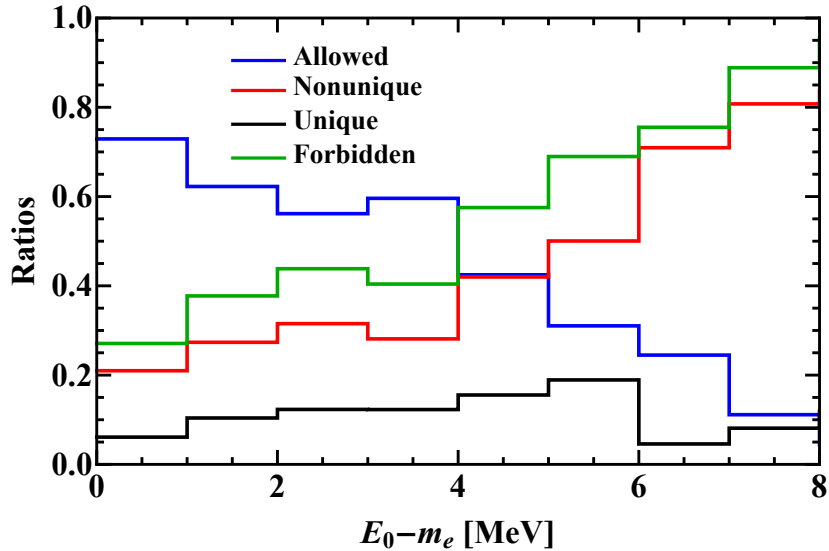}
	\caption{The relative ratios of different beta decay transitions as the function the endpoint energy for the $^{235} {\rm U}$ fission isotope.}
\end{figure}

Next we need the relation between the nuclear charge number $Z$ of the daughter nucleus and the endpoint energy $E_0-m_{e}$ which plays a significant role in
the effective conversion calculation~\cite{Vogel:2007du}. The distributions of $Z$ with respect to $E_0-m_{e}$ are shown in Fig.~3
for the fission isotope $^{235} {\rm U}$, where in the left panel all the decay branches are assumed to be the allowed GT transition, and in the right panel
the results for the allowed and forbidden transitions are illustrated separately. Here we combine the unique and nonunique forbidden transitions
for simplicity. In each energy bin, the effective nuclear charge number $\overline{Z}$ is calculated as~\cite{Huber:2011wv,Vogel:2007du}
\begin{align}
\overline{Z}(E_0) &= \frac{\sum_{A,Z} Y(A,Z-1) \sum_{i} b_i(E^i_0) Z}{ \sum_{A,Z} Y(A,Z-1) \sum_{i} b_i(E^i_0)} \;,
\end{align}
in which $Y$ is the fission yield corresponding to the mother nucleus of the beta decay with $A$ and $Z-1$,
$E^i_0$ is the total energy release in the decay branch, and then $E^i_0-m_e$ is the corresponding endpoint energy,
$b_i$ is the branching ratio of the $i$-th branch for the nucleus with $E^i_0$ in the selected energy bin of $E_0$.
We use a second-order polynomial to fit the effective nuclear charge number as the function of the endpoint energy
\begin{eqnarray}
\overline{Z}(E_0) = a_{0} + a_{1} (E_0-m_{e}) + a_{2} (E_0-m_{e})^{2} \;,
\end{eqnarray}
where fitting curves are shown in the left and right panels of Fig.~3, 
and the corresponding coefficients are summarized in Tab.~2. Note that for a nuclear charge number
the forbidden transition tends to have larger endpoint energy compared to the allowed transition. It seems that coefficients for the case of taking all the branches as allowed are rather different from those in Refs.~\cite{Vogel:2007du,Huber:2011wv}. In our study, $\overline{Z}$ is the charge of the daughter nucleus, while in Refs.~\cite{Vogel:2007du,Huber:2011wv} it is the charge of the mother nucleus. Considering this difference, our $\overline{Z}(E_0)$ curve agrees with that in Ref.~\cite{Huber:2011wv} within 0.5 unit of $\overline{Z}$ above 2 MeV, which mainly comes from different versions of the nuclear database.
In the next two subsections, we shall present two different realizations of the conversion calculations.
In Sec.~3.2, the conversion calculation assumes all the virtual branches are the allowed GT transition,
thus the fitting curve in the left panel of Fig.~3 will be employed.
This realization has been widely used in previous calculations~\cite{Vogel:2007du,Huber:2011wv}
but its reliability is challenged by the recent observed reactor rate and spectral
anomalies~\cite{Mention:2011rk,Giunti:2019qlt,Seo:2014xei,An:2015nua,Abe:2015rcp,Ko:2016owz}.
Thus a new realization of the conversion calculation will be presented in Sec.~3.3 considering
the energy spectra of virtual branches from both the allowed and forbidden beta decays, in which
the $\overline{Z}(E_0)$ and $E_0-m_{e}$ relations in the right panel of Fig.~3 will be used.
\begin{figure}
\centering
\includegraphics[width=0.45\linewidth]{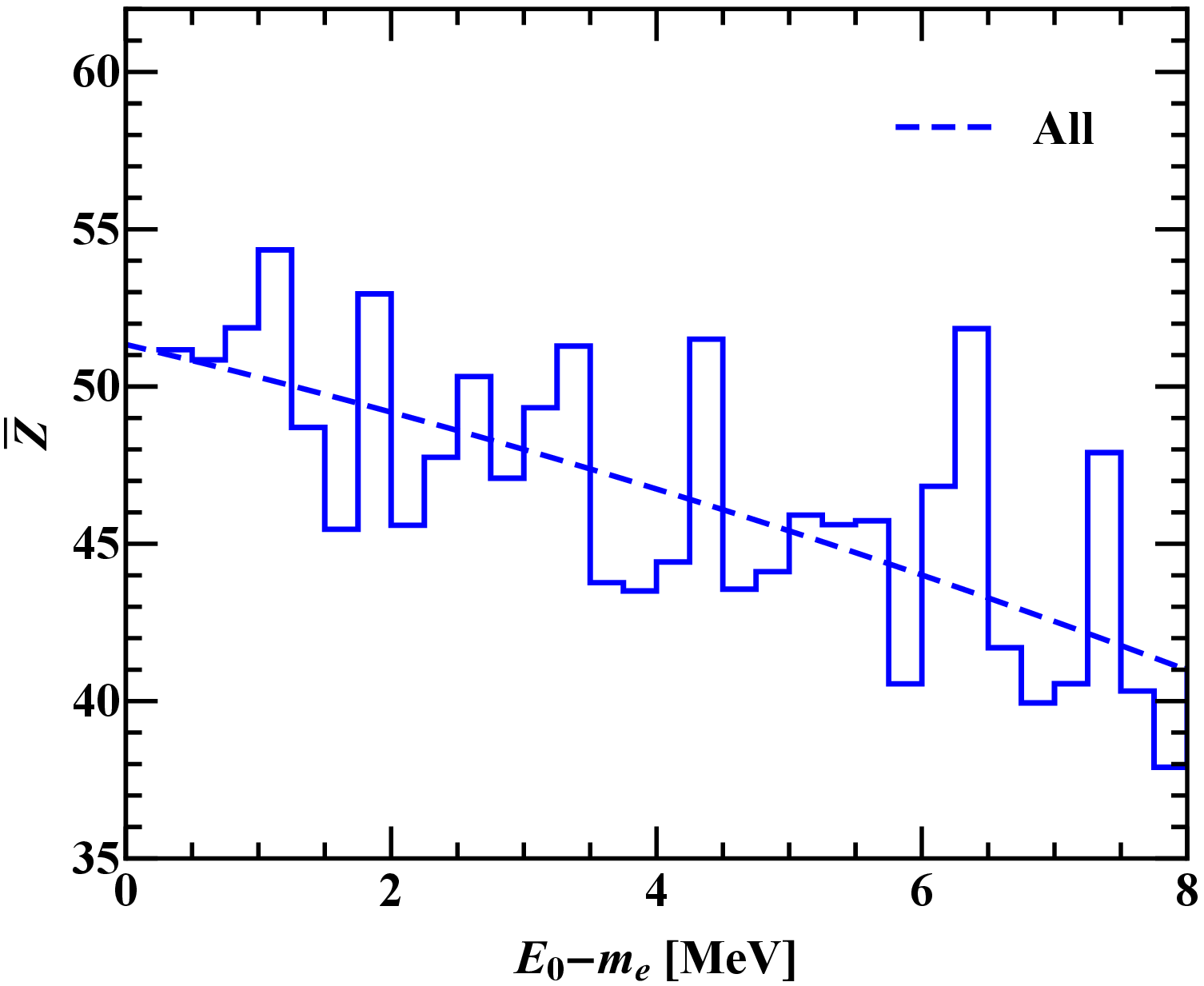}
\includegraphics[width=0.45\linewidth]{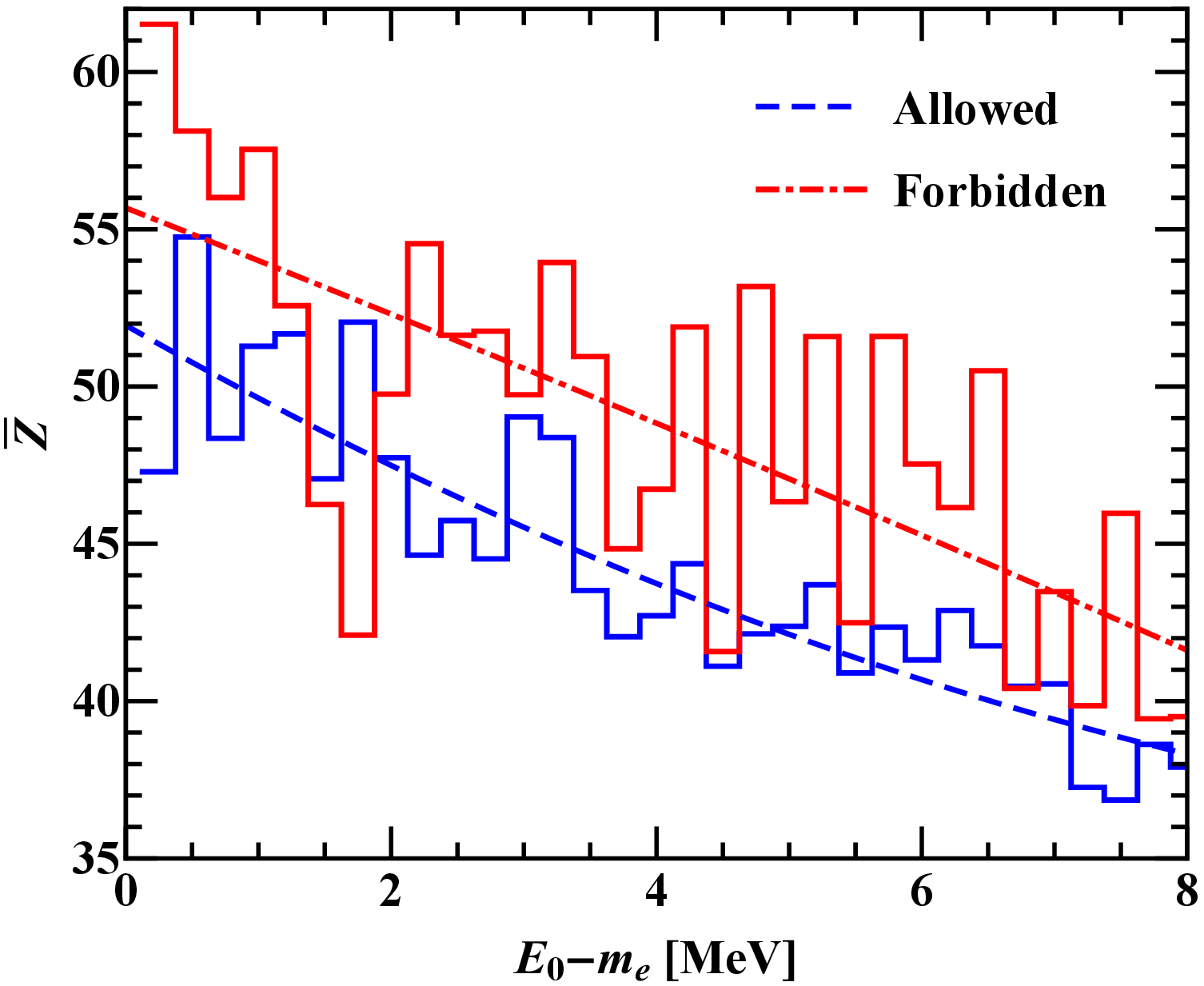}
\caption{The relation between the effective nuclear charge number $\overline{Z}$ and endpoint energy $E_0-m_{e}$ for the $^{235}\rm{U}$ fission isotope.
All the decay branches are assumed to be the allowed GT transition in the left panel,
and the results for the allowed and forbidden transitions are illustrated separately in the right panel.
The second-order polynomial curves are fitted as the dashed or dashed and dash-dotted lines in the left and right panels.}
\end{figure}
\begin{table}[h!]
\centering
	\caption{The coefficients for the polynomial fit of the effective nuclear charge number as the function of the endpoint energy $E_0-m_{e}$.}
	\vspace{0.25cm}
	\begin{tabular*}{\hsize}{p{0.1\textwidth}p{0.3\textwidth}<{\centering}p{0.3\textwidth}<{\centering}p{0.3\textwidth}<{\centering}} 
		\hline
		\hline
		 & $a_0$ & $a_1$ (MeV$^{-1}$) & $a_2$ (MeV$^{-2}$)
		\\
		\hline
		All & 51.3374 & -1.00324 & -0.0363509
		\\
		Allowed & 51.9464 & -2.40159 & 0.0873305
		\\
		Forbidden & 55.6795 & -1.66458 & -0.0116643
	    \\
		\hline
		\hline
	 \end{tabular*}
\end{table}

\subsection{Conversion with allowed virtual branches}

In the conversion calculation of the isotopic antineutrino flux of $^{235}\rm{U}$, we need the \textit{ab initio} calculations of the
isotopic electron and antineutrino fluxes, and treat the electron spectrum as the mock data to make the conversion calculation. The electron and antineutrino spectra are generated in 50 \rm{keV} bins in the energy range from 0 to 10 \rm{MeV}, namely, 200 data points for one single spectrum. The number of virtual branches is determined by the number of total data points and the number of data points for each branch. We have compared the performance of different grouping methods and demonstrated that four data points are the optimal choice for the whole region of the electron spectrum. Then 50 virtual beta decay branches of the allowed GT transition will be used to convert the electron spectrum to the antineutrino spectrum.
To test the reliability and accuracy of the conversion procedure, we use four kinds of treatments regarding the forbiddenness of the decay branches
in the \textit{ab initio} calculations. First we follow the assumption in Ref.~\cite{Vogel:2007du,Huber:2011wv} and take all the branches to be allowed GT transition (case A),
Moreover, both the allowed and forbidden transitions will be considered in the \textit{ab initio} calculations,
assuming the nonunique forbidden decays belong to the GT $0^{-}$ transition (case B) or the GT $1^{-}$ transition (case C)
or a mixture of GT $0^{-}$ and $1^{-}$ transition (case D).
Meanwhile, the unique forbidden decays are taken as the GT $2^{-}$ transition for all the three conversion scenarios.

The conversion procedure is briefly described as follows.
We first divide the whole energy range of the electron spectrum into 50 equal-size energy intervals.
Starting from the interval of highest energies in the electron energy spectrum,
we fit the electron spectrum in this window using a description of the beta decay according to Eq.~(1)
including two free parameters, namely the normalization and endpoint energy. The nuclear charge number is calculated using the
second-order polynomial in the left panel of Fig.~3. The antineutrino energy spectrum in this window is obtained
by the substitution of $E_{\overline{\nu}} = E_0 - E_e$ and a replacement of the radiative corrections from the electron to antineutrino one as in Eq.~(4).
Next we are going to move to the second energy interval next to the first one.
After the best-fit electron spectrum in the previous energy window is properly removed,
a fit to the electron spectrum in this window is achieved using a new beta decay shape function with two free parameters.
Then the same procedure is repeated for all the intervals in the order from the high to low energies.
Finally the fitted electron and converted antineutrino spectra
are obtained with a direct summation of the individual electron or antineutrino spectrum of each virtual beta branch.
To suppress the spurious fluctuation, a rebinning average process from the 50 keV bins to 250 keV bins has been applied to the converted antineutrino spectrum.

\begin{figure}
\centering\includegraphics[width=0.8\linewidth]{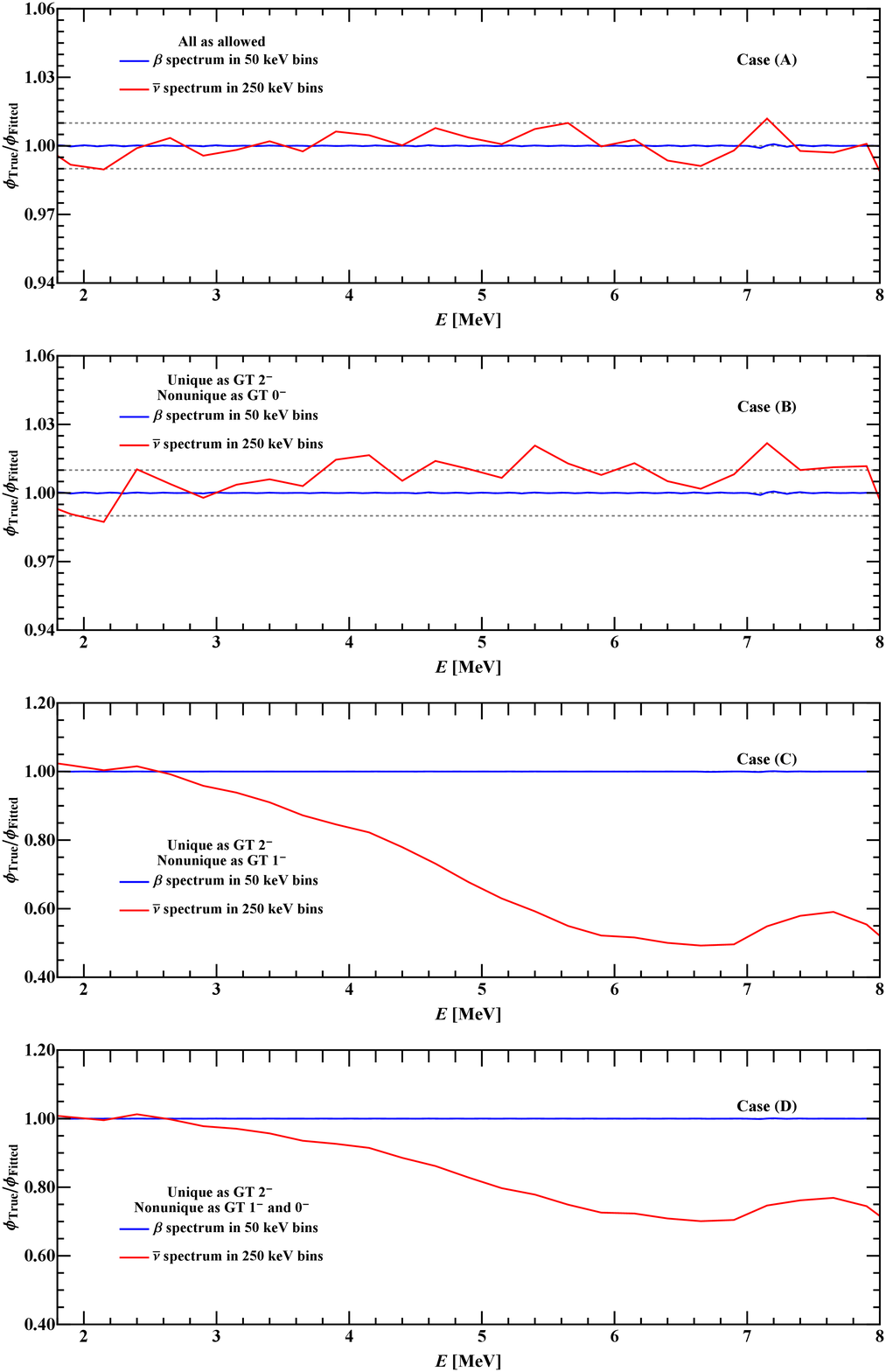}
\caption{The ratios of the true electron or antineutrino spectra of the \textit{ab initio} calculations to the corresponding converted spectra for $^{235}{\rm U}$,
where case A assumes all the branches are the allowed GT transition (first panel), and case B, C, D assume that the nonunique forbidden decays are subject
to the GT $0^{-}$ transition (second panel), the GT $1^{-}$ transition (third panel) or a mixture of GT $0^{-}$ and $1^{-}$ transitions (fourth panel).
The unique forbidden decays are taken as the GT $2^{-}$ transition for the last three cases.}
\end{figure}
To test the degrees of consistency in the conversion method, the ratios of the true electron or antineutrino spectra of the
\textit{ab initio} calculations to the corresponding converted spectra are calculated and illustrated in Fig.~4,
where for all the cases the converted electron spectra are in perfect agreement with the \textit{ab initio} calculations,
and thus demonstrating the excellent fitting quality for all the conversions. Note we only show results in the energy range from 1.8 to 8 \rm{MeV}, which is relevant to the current measurements of antineutrino spectra.

Next let us focus on the converted antineutrino spectra.
For case A when the \textit{ab initio} calculation is constructed by assuming all the decay branches are the allowed GT transition,
a reliable antineutrino spectrum has been achieved with the accuracy better than 1\% in the whole energy range from 1.8 to 8 MeV
in the first panel of Fig.~4, which turns out to be consistent with the conclusion in Ref.~\cite{Vogel:2007du,Huber:2011wv}.
In the realistic calculation there are as many as 30\% of the decay branches being different types of forbidden transitions,
the bias induced by the misdescription of beta decay spectra in the allowed virtual branches of the conversions are shown
in the second, third and fourth panels of Fig.~4. We observe that in general the bias is negligible below 3 MeV and start to
increase as the energy grows, but the sizes and shape features are very different for case B, C, and D. The deviation is around 2\%
for case B, and can reach 50\% and 30\% for case C and D, where the GT $1^{-}$ transition contributes to 100\% and 50\% of
the nonunique forbidden decays.
We can have a more clear view on the effects of different forbidden transitions by combining the unique and nonunique
forbidden decays and assuming the pure GT $0^{-}$, $1^{-}$, or $2^{-}$ transition for all the forbidden decays, where the results are respectively
shown as the black curves in the upper, middle and lower panels of Fig.~5. It is obvious that the largest deviations are 1\%, 55\% and
5\% for the GT $0^{-}$, $1^{-}$, and $2^{-}$ transitions, respectively.
One should be noted that the actual size of the shape deviation strongly depend on the percentage and ratios of different forbidden transitions.
Qualitatively speaking, our conclusions here are consistent with the conclusion in Ref.~\cite{Hayes:2013wra},
but our analysis is self-consistent and based on the state-of-the-art nuclear database.

The very distinct effects of different forbidden transitions can be
understood using their spectral characteristics shown in Fig.~1. The shape difference between the GT allowed and forbidden $0^{-}$ transitions is
only at the 5\% level and thus it is reasonable to have a 1\% deviation in the converted antineutrino spectrum. The shape variation is dramatic for the GT $1^{-}$, and $2^{-}$ transitions but their effects on converted antineutrino spectra are very different.
The reason lies in the shape of spectral ratios in Fig.~1, where it is similar to parabolic curve for the GT $2^{-}$ transition but it is a
monotonically decreasing function for the GT $1^{-}$ transition. For the former case, the shape variation is localized and when fitted
to the electron spectrum the bias can be largely reduced by tuning the normalization factors of virtual beta decays. On the other hand,
it is not possible to significantly compensate the bias of the shape variation for the latter case since the monotonic shape variation
of each virtual branch will have divergent contributions to the whole energy range.

\subsection{Conversion with both allowed and forbidden virtual branches}

In the conversion calculation of the isotopic antineutrino fluxes, neglecting shape variation of beta decay branches between
the allowed and forbidden transitions would induce significant bias in the total inverse-beta-decay yields and energy spectral
distributions. Therefore, it is very important to take account of the proper contribution from forbidden transitions. However, since the
virtual branch does not correspond to a real physical beta decay, one cannot assign appropriate quantum numbers and
physical parameters to the initial and final nucleus states for one particular virtual decay branch. In order to obtain suitable nuclear
charge numbers and shape factors for the virtual branches, we rely on the statistical information of the state-of-the-art
nuclear database.

In principle, we need to distinguish the allowed and all kinds of forbidden transitions, and
to calculate the relative ratios of all the transition types as functions of the endpoint energy,
and also the nuclear charge number of each type as the function of the endpoint energy.
In the current work, since available information of the latest nuclear database
is still not complete and accurate enough, we combine all the forbidden transitions
and are going to demonstrate the viability of the new realization of the conversion method
using only two kinds of beta decay modes: the allowed and forbidden transitions.
Therefore, the relation between the relative ratios of allowed and forbidden transitions
and the endpoint energy, and that of nuclear charge numbers versus the endpoint energy
are shown in Fig.~2 and the right panel of Fig.~3, respectively.


The conversion procedure in the new realization is rather similar to the one in the previous subsection,
so here we only highlight the main differences of this new method. First, after selecting an energy
interval of the electron spectrum in each conversion step, we have to decide whether the virtual decay branch is allowed or
forbidden according to the probability distributions of the relative ratios as shown in Fig.~2. Then the fitting
is done with a proper shape function of the allowed or forbidden transition
using the correct shape factor and correct relation between the nuclear charge number and endpoint energy.
Second, because there are only 50 virtual decay branches for each time of the conversion, the sampling fluctuation would
be large for the relation between the relative ratios and endpoint energy,
so in the conversion, we repeat 100 independent conversion calculations and take
the average of the converted antineutrino spectra
as the final antineutrino spectrum of $^{235} {\rm U}$.
Finally since we are not going to deal with the mixture of three different kinds of forbidden transitions,
three extreme cases that all the forbidden decays are respectively treated as the GT $0^{-}$, $1^{-}$, or $2^{-}$ transition are
considered to reveal the advantages of the new conversion method.
\begin{figure}[!htb]
\centering\includegraphics[width=0.9\linewidth]{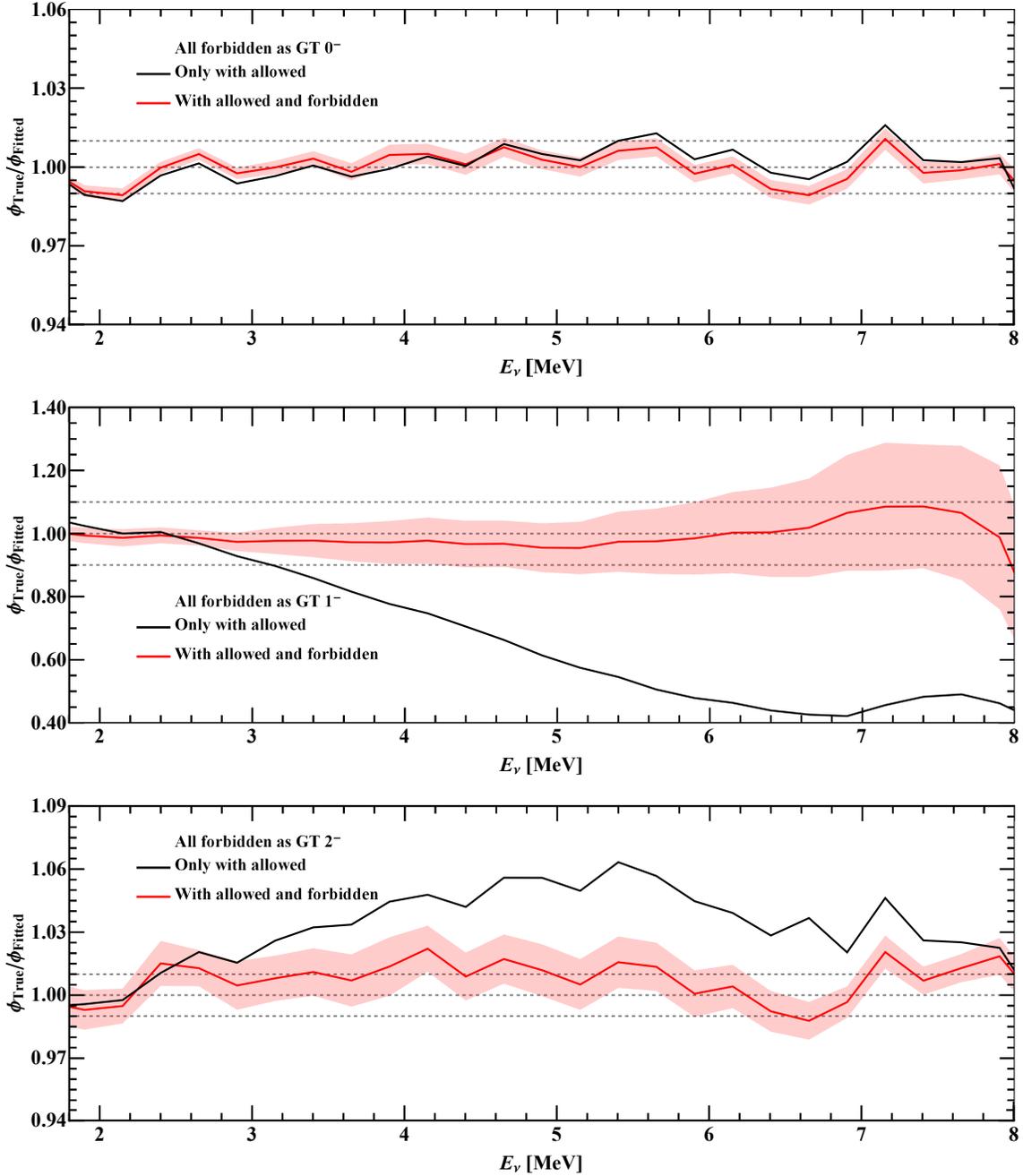}
\caption{
The ratios of the true antineutrino spectra of the \textit{ab initio} calculations to the corresponding converted spectra for $^{235}{\rm U}$,
where all the forbidden transitions are treated as the GT $0^{-}$ (upper panel), $1^{-}$ (middle panel), and $2^{-}$ (lower panel) transitions respectively.
The cases of conversions with all allowed virtual branches are shown in black lines and the cases of conversions with both allowed and forbidden virtual branches are
shown in red lines. The red shadowed bands are shown as the standard deviations of 100 converted antineutrino spectra.}
\end{figure}

The ratios of the true antineutrino spectra of the \textit{ab initio} calculations to the corresponding converted spectra for $^{235}{\rm U}$ in the
new conversion method are illustrated in Fig.~5, where all the forbidden transitions are treated as the GT $0^{-}$ (upper panel), $1^{-}$ (middle panel),
and $2^{-}$ (lower panel) transitions respectively.
The red shadowed bands are shown as the standard deviations of 100 converted antineutrino spectra and can be regarded as
one source of the flux uncertainties.
Compared to the conversion with only allowed virtual branches, a consideration of both
allowed and forbidden virtual branches does significantly improve the degrees of agreement, in particular for the GT $1^{-}$, and $2^{-}$ transitions.
There is also a moderate improvement for the GT $0^{-}$ transition, and now the degree of agreement is better than 1\%.
For the GT $2^{-}$ transition, the deviation is largely reduced to the level of 1\% in all the energy range,
and neglecting the forbidden transition would induce an excess in the high antineutrino energy range.
For the GT $1^{-}$ transition, it is remarkable to note that
the maximal deviation of around 40\% to 50\% can be reduced to the level of 5\% for most of the antineutrino energies.
Regarding the uncertainties, we observe that the band widths are correlated with the spectral deviations
of neglecting forbidden transitions.
These uncertainties are around 1\% and 3\% for the GT $0^{-}$ and $2^{-}$ transitions respectively, but for the case of
the GT $1^-$ transitions it grows from around 2\% at 2 MeV to larger than 10\% at above 6 MeV, which can be understood
by the very distinct spectral features between allowed and GT $1^-$ beta decays and the limited numbers of the virtual branches.
Since here we only consider the extreme case, if the GT $1^{-}$ transition should contribute to 20\% of the forbidden decays,
the deviation in the current realization of conversion calculations would be at the level of better than 1\%
and the induced uncertainty would be at the level of 2\% at around 6 MeV.
As we have explained in the previous subsection, the shape difference between the allowed and GT $1^{-}$ forbidden
transitions tends to produce a large deviation in the converted antineutrino spectrum. Thus there is still
room for the further improvement regarding the distributions for the ratio of forbiddenness and nuclear charge number
in the high endpoint energy range because of low statistics and large fluctuations of these distributions at the
high endpoint energies.
\begin{figure}[!htb]
\centering\includegraphics[width=0.9\linewidth]{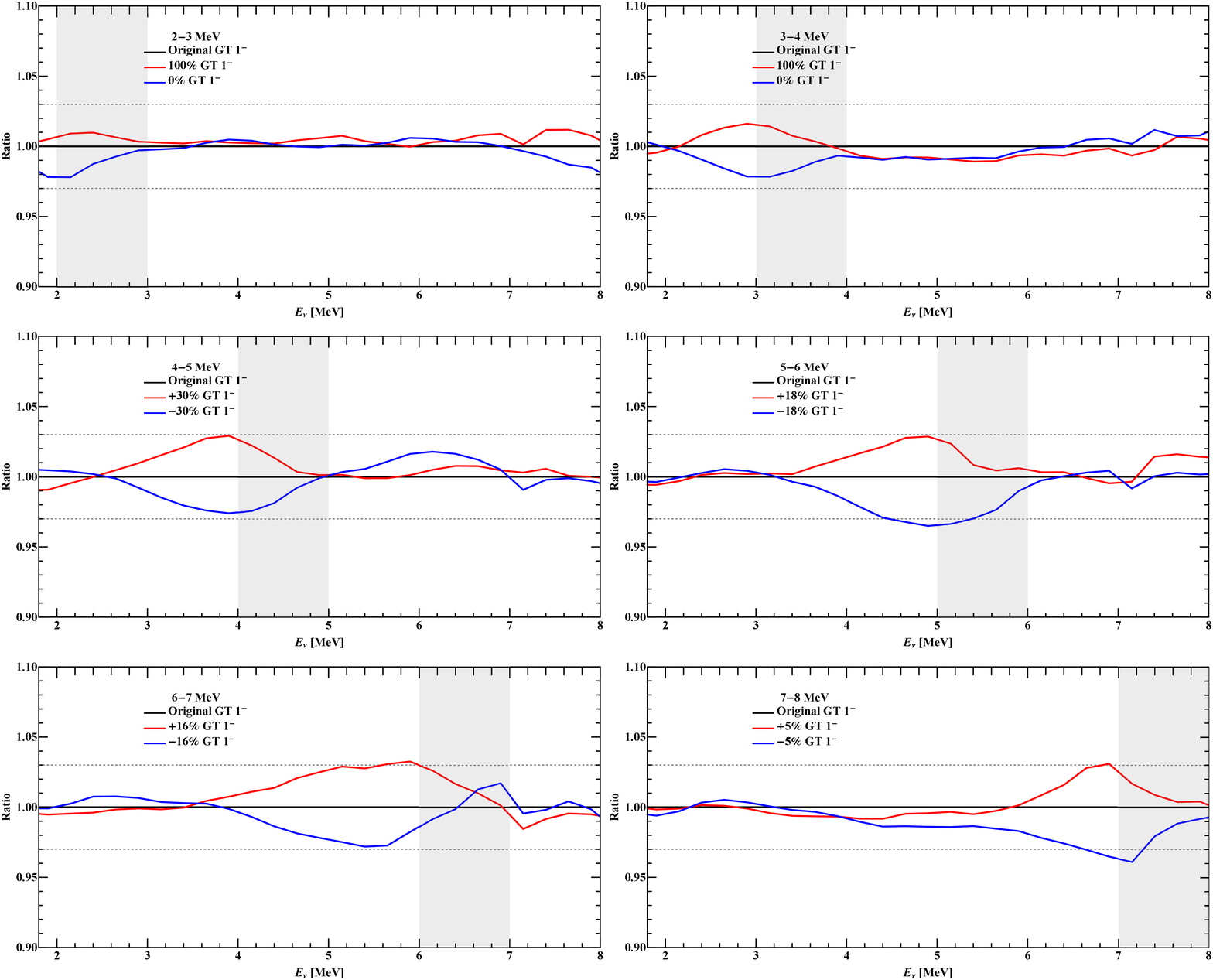}
\caption{
The ratios of the converted $^{235}{\rm U}$ antineutrino spectra
after and before changing the relative ratios in the selected endpoint energy windows,
where a benchmark requirement of the spectral deviation better than 3\% is assumed.
The contributions of the allowed and GT $2^{-}$ transitions are fixed to their true values of Fig.~2.}
\end{figure}

To illustrate the requirement on the ratios of forbiddenness, in particular for those between
GT $0^{-}$ and $1^{-}$ transitions which are not properly provided in the current nuclear database,
we make a sensitivity study on how the converted antineutrino spectra depend on
the relative ratio (from 0 to 100\%) between the GT $0^{-}$ and $1^{-}$ transitions.
Fig.~6 shows the ratios of the converted $^{235}{\rm U}$ antineutrino spectra
after and before changing the relative ratios in the selected endpoint energy windows,
where a benchmark requirement of the spectral deviation better than 3\% is assumed.
Each panel from the upper left to the lower right ones corresponds to the
modification of the relative ratio from the low to high endpoint energy window respectively.
The contributions of the allowed and GT $2^{-}$ transitions are fixed to their true values of Fig.~2.
From the figure, one can observe that, to achieve an accuracy of better than 3\%, the relative ratio
between the GT $0^{-}$ and $1^{-}$ transitions should
be controlled within the precision of 5\%, 16\%, 18\% and 30\% for the endpoint energy windows of
[$i,\,i+1$] MeV ($i=7, 6, 5, 4$) respectively. Meanwhile, the relative ratios between the GT $0^{-}$ and $1^{-}$
transitions can be arbitrary under the condition of better than 3\% for the endpoint energies below 4 MeV,
where the allowed transitions predominately contributes.


To summarize, one should note that there are two groups of main uncertainties in the conversion process.
The first one is the variation of different samplings due to limited virtual branches which can be improved
with more accurate beta spectra and much more numbers of virtual branches. The second one is
the relative ratios of the allowed and different forbidden transitions as functions of the endpoint energies,
and a reliable classification of the allowed and forbidden transitions in the nuclear database is required.
To achieve this goal, a careful survey on the beta decay properties of individual decay branch will be mandatory~\cite{Rice:2017kfj,Guadilla:2019aiq},
and those with high fission yields, high branching ratios and high endpoint energies share the high priority for the realistic studies.

\section{Application to the ILL beta spectra}
As revealed in the study using the mock data from nuclear database, considering the forbidden transition in the conversion of virtual beta branches can remove the significant bias and produce a reliable isotopic antineutrino spectrum as long as the statistical distributions of different forbidden transitions are accurately known. However, it is not the case for current real situation of the isotopic fission products.
The unknown information on the nonuniqe forbidden GT $0^-$ and $1^-$ transitions may induce large variations to the converted antineutrino spectra and constitute a
large source of the conversion uncertainty.
Similar to Sec.~3.3, in this section we apply the same conversion procedure to the measured beta spectra at ILL. We take the $^{235}{\rm U}$ beta spectrum from Ref.~\cite{Haag:2014kia} to illustrate the conversion properties. Similar results have been obtained for the other two isotopes ${}^{239}\text{Pu}$,
${}^{241}\text{Pu}$, and the conclusion remains unchanged.

In the following calculation, after optimization by comparing different choices,
we employ 24 virtual branches in the energy range from 1.5 to 8.65 MeV.
We first make the conversion assuming all the virtual branches are allowed, and the
conversion results of the ILL $^{235}{\rm U}$ beta spectrum are shown in Fig.~7.
The ratio is defined as the fitted spectrum to the ILL data for electrons,
whereas it is the ratio of the converted spectrum to the model prediction of Ref.~\cite{Huber:2011wv}
for antineutrinos. From the figure one can conclude that the fitted electron and antineutrino spectra
with all allowed virtual branches show excellent agreements with ILL data and previous model prediction
within the 1\% and 2\% ranges for most of the energies from 1.8 to 8 MeV. These small differences between antineutrino spectra can be explained by slightly different $\overline{Z}(E_0)$ relation and different forms of corrections.
\begin{figure}
\centering\includegraphics[width=0.9\linewidth]{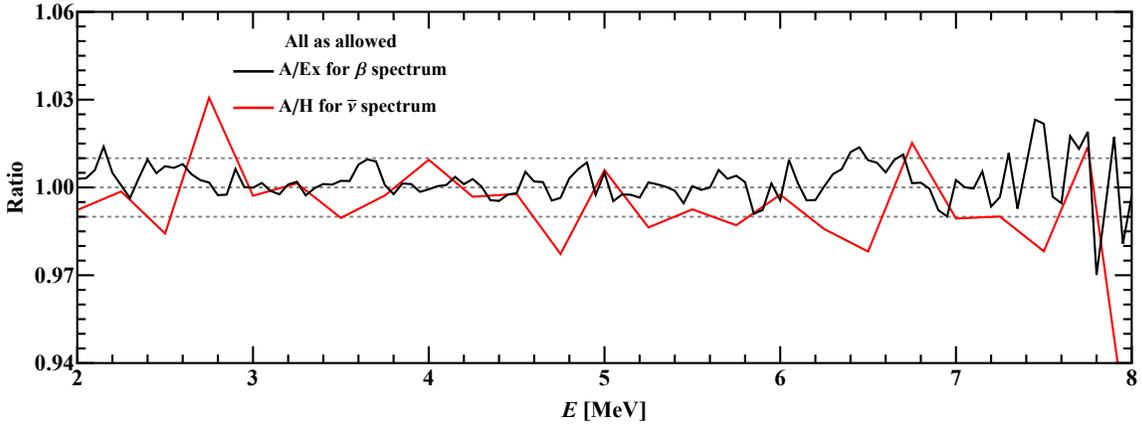}
\caption{
Conversion results of the ILL $^{235}{\rm U}$ beta spectrum using allowed GT transitions.
The ratio is defined as the fitted spectrum to the ILL data for electrons,
whereas it is the ratio of the fitted spectrum to the model prediction of Ref.~\cite{Huber:2011wv}
for antineutrinos.}
\end{figure}

\begin{figure}[!htb]
\centering\includegraphics[width=0.8\linewidth]{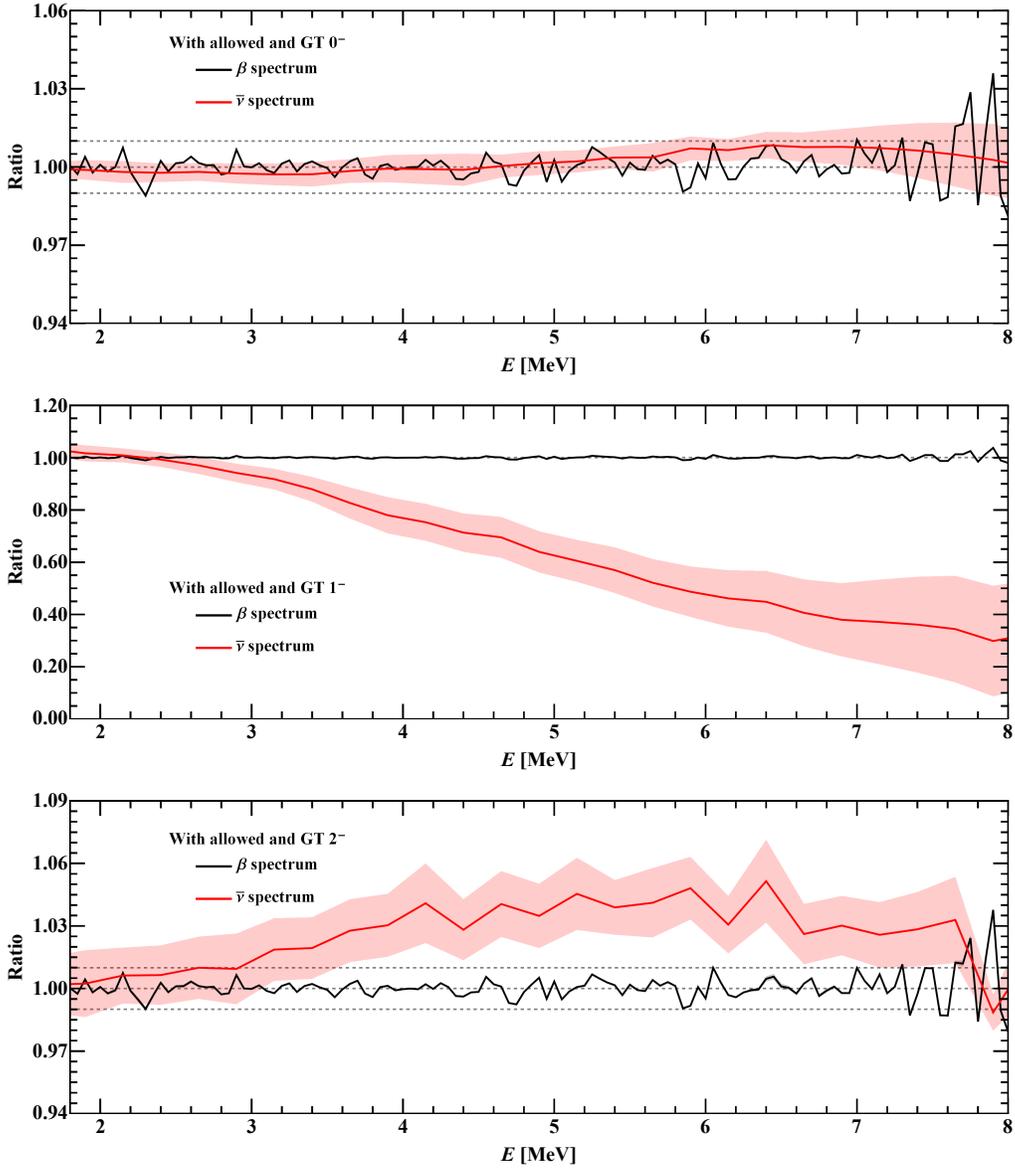}
\caption{
Conversion results of the ILL $^{235}{\rm U}$ beta spectrum using both allowed and forbidden transitions.
The first, second, third and panels are the results for the GT $0^-$, $1^-$, $2^-$ respectively, and the ratio is defined as the fitted spectrum to the ILL data for electrons, and as the converted spectrum using both allowed and forbidden virtual transitions to the spectrum using only the allowed transition for antineutrinos. The red shadowed bands are shown as the standard deviations of 100 converted antineutrino spectra.}
\end{figure}

Next we want to realize the conversion calculation of the ILL $^{235}{\rm U}$ beta spectrum using both the allowed and the forbidden virtual transitions.
For the classification of the forbidden transitions, we consider three extreme scenarios of assuming all the forbidden transitions are the GT $0^-$, $1^-$ or $2^-$ type respectively. We employ the probability ratio distributions of the allowed and forbidden transitions in Fig.~2 to determine the types of virtual beta branches.
Similar to the calculation as in Fig.~5, for each scenario we repeat the conversion of the ILL $^{235}{\rm U}$ beta spectrum 100 times, and take their average as the converted antineutrino spectrum and the standard deviation as the induced uncertainty by finite numbers of virtual branches, which are illustrated in Fig.~8,
where the first, second, third panels are the results for GT $0^-$, $1^-$, $2^-$ respectively, and the ratio is defined as the fitted spectrum to the ILL data for electrons, and as the converted spectrum using both allowed and forbidden virtual transitions to the spectrum using only the allowed transition for antineutrinos.

From Fig.~8, we can learn that the ILL electron spectrum is fitted very well for all three cases and the behavior of the converted antineutrino spectra agrees with the results in Fig.~5. For the GT $0^-$ and GT $2^-$ transitions, the induced spectral variations are about 1\% and 5\% respectively, which are consistent with the evaluation in Ref.~\cite{Hayes:2013wra} that the uncertainty induced by the inclusion of first forbidden transitions is about 4\%.
The GT $1^-$ transition brings the largest spectral variation into the conversion procedure which is much larger than the level of 4\%.
This is mainly due to the shape factor difference of the GT $1^-$ transition between the case of PWA used in Ref.~\cite{Hayes:2013wra} and the case of ERC in the current manuscript. One can refer to Tab.~1 and the middle panel of Fig.~1 for the expressions and energy-dependent behavior of two kinds of shape factors.

\begin{figure}
\centering\includegraphics[width=0.66\linewidth]{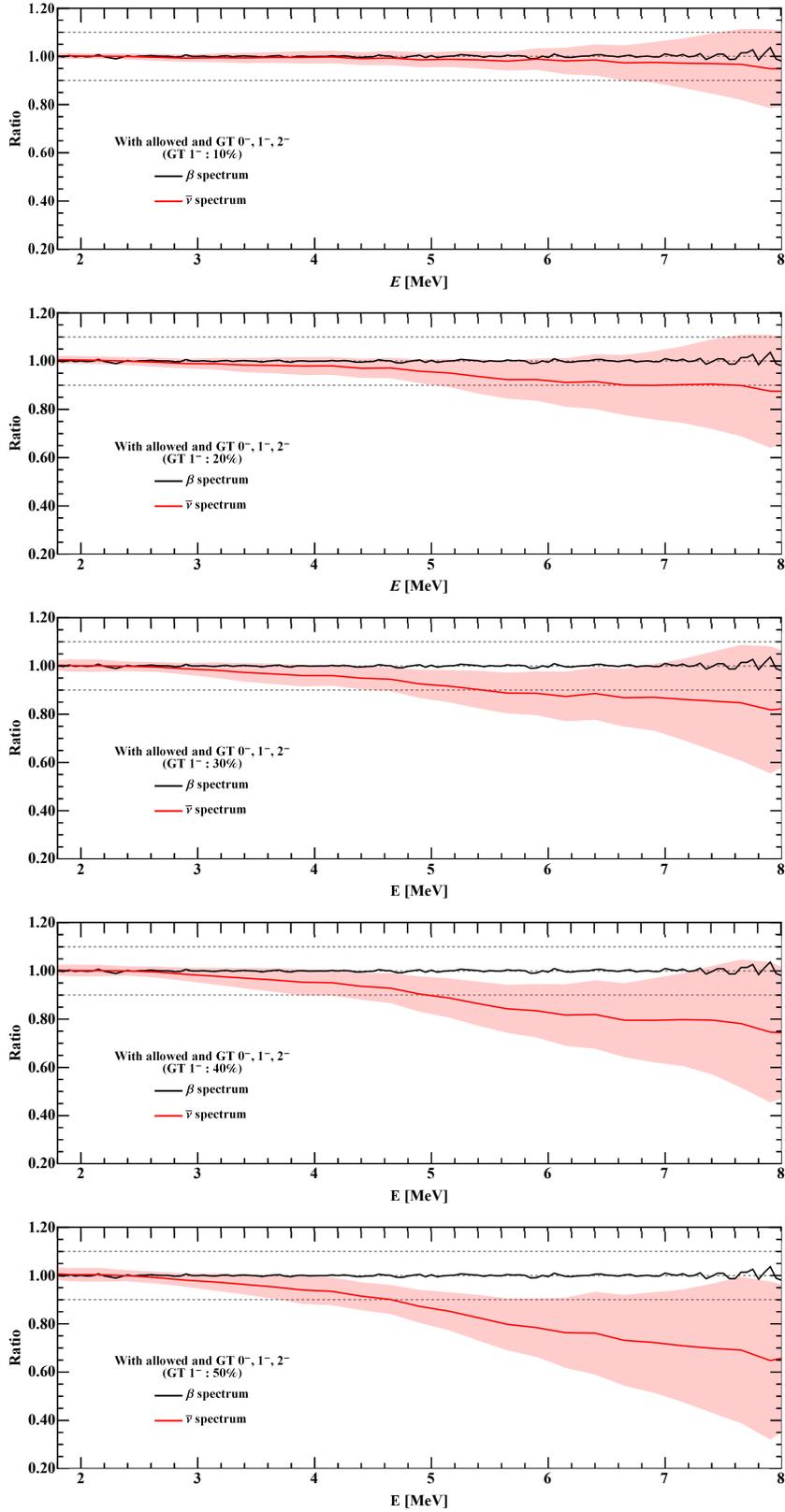}
\caption{
Conversion results of the ILL $^{235}{\rm U}$ beta spectrum using both allowed and all types of GT $0^-$, $1^-$, $2^-$ forbidden transitions.
the ratio is defined as the fitted spectrum to the ILL data for electrons, and as the converted spectrum using both allowed and forbidden virtual transitions to the spectrum using only the allowed transition for antineutrinos. The red shadowed bands are shown as the standard deviations of 100 converted antineutrino spectra.
}
\end{figure}

Considering the realistic case as shown in Fig.~2, we are going to encounter a mixture scenario where
GT $0^-$, $1^-$ and $2^-$ transitions are all involved. To realize a conversion calculation including
all the forbidden transitions, we take different relative ratios between the GT $0^-$ and $1^-$ transitions
and the contributions of the allowed and GT $2^-$ transitions are assumed to be fixed as in Fig.~2.
We illustrate the conversion results of the mixture scenarios in Fig.~9, where the first to fifth panels correspond to the
cases that the GT $1^-$ transition contributes to 10\%, 20\%, 30\%, 40\%, 50\% of the nonunique beta decays.
We repeat 100 independent conversion calculations and take the average and standard deviation as the converted antineutrino
spectrum and the corresponding uncertainty. As the relative ratio of the GT $1^-$ transition deceases, the spectral deviation
and uncertainty band are both reduced accordingly. From the current reactor antineutrino measurements, it might be reasonable to
take a 10\% spectral deviation from the allowed-only conversion as the limit for model predictions. In this circumstance,
one may estimate that the contribution
of the GT $1^-$ transition should be less than around 20\% in the total nonunique transitions.

Therefore, in order to obtain accurate isotopic antineutrino spectra in the conversion calculation,
a prerequisite is to get accurate statistical information on the beta decay branches that contribute significantly to
the isotopic fission process, in particular for the most important GT $1^-$ transition.
This requires a careful summary and evaluation of the beta decay and fission yield information from
different nuclear database and theoretical calculations which is beyond the scope of the current work and will the studied elsewhere.

\section{Concluding remarks}

Predicting the isotopic antineutrino fluxes from ${}^{235}\text{U}$,
${}^{238}\text{U}$,
${}^{239}\text{Pu}$, and
${}^{241}\text{Pu}$ has been always an important task for reactor antineutrino experiments.
In general there are two categories of predicting methods, where the first one is
the \textit{ab initio} summation method with the nuclear database, and the second one is the effective conversion
method based on the measurements of aggregate electron spectra associated with fission isotopes. However,
the appearance of reactor antineutrino flux and spectral anomalies in current reactor antineutrino experiments
and severe requirement for the neutrino mass hierarchy measurement in future reactor antineutrino experiments
have challenged the current model predictions of the reactor antineutrino fluxes, and new reliable calculations
are required to resolve the reactor anomalies and serve as standard inputs for future measurements.

In this work we have examined the reliability and accuracy of the conversion method using the \textit{ab initio}
calculations of the electron and antineutrino spectra by means of the state-of-the-art nuclear database. Furthermore,
we have proposed a new realization of the conversion calculation with both the allowed and forbidden virtual branches
by virtue of the statistical properties of the allowed and forbidden decays in the nuclear database.
Applications to both the \textit{simulated} data of the electron spectrum from the nuclear database and the \textit{real}
data from the fission measurement at ILL are also presented, and large spectral variation has been observed
due to different assumptions of the forbidden virtual branches.
We have observed that neglecting the shape variation of beta decay branches between the allowed and forbidden
transitions would induce significant bias in the total inverse-beta-decay yields and energy spectral distributions,
among which the GT $1^{-}$ forbidden transition has the largest effects because of the monotonically decreasing shape
in the energy spectral ratio of the forbidden and allowed transitions.
Two kinds of dominant uncertainty sources are identified and
it has been proved that the new conversion method can reduce the rate and spectral bias and present a
reliable prediction of the antineutrino fluxes
as long as we have accurate measurements of the isotopic electron energy spectra and reliable statistical information
on the relative ratios and nuclear charge numbers for the selected classification of the allowed and forbidden decay
transitions. Finally a more specific application of this new conversion method to the ILL data of aggregate electron spectra
and a complete calculation of the uncertainty associated with the conversion method require careful evaluations of statistical properties of
beta decay branches, which will be presented in a separated work in the near future.

\section*{Acknowledgements}
The authors would like to thank Dongliang Fang, Carlo Giunti, Zhi-zhong Xing for helpful discussions.
This work was supported in part by the National Natural Science Foundation of China under Grant No.~11835013 and No.~11775231,
by the Strategic Priority Research Program of the Chinese Academy of Sciences under Grant No. XDA10010100,
and by the CAS Center for Excellence in Particle Physics (CCEPP).


\end{document}